\documentclass[12pt,letter]{emulateapj}

\newcommand{\idrop}{$i_{775}$-dropout } 
\newcommand{\idrops}{$i_{775}$-dropouts }
\newcommand{\vdrop}{$V_{606}$-dropout } 
\newcommand{\iz}{$i_{775} - z_{850}$}
\newcommand{\z}{$z$ } 
\newcommand{\m}{$m$ } 
\newcommand{\M}{$M$ }
\newcommand{\six}{$z\sim6$ } 
\newcommand{\is}{$i_{775}$} 
\newcommand{\zs}{$z_{850}$}
\newcommand{\f}{$f$-factor } 
\newcommand{\err}[2]{^{+#1}_{-#2}}

\shorttitle{UDF05 FOLLOW-UP OF HUDF. III.}
\shortauthors{Su et al.}

\begin{document} 
\title{The UDF05 Follow-up of the Hubble Ultra Deep Field. III. The Luminosity Function at \six \altaffilmark{1}}

\altaffiltext{1}{Based on observations made with the NASA/ESA Hubble Space Telescope, obtained at the Space Telescope Science Institute, which is operated by the Association of Universities for Research in Astronomy, Inc., under NASA contract NAS 5-26555. These observations are associated with program \#10632 and \#11563.}

\author{
Jian Su\altaffilmark{2,3},
Massimo Stiavelli\altaffilmark{3}, 
Pascal Oesch\altaffilmark{4,5,6}, 
Michele Trenti\altaffilmark{7},
Eddie Bergeron\altaffilmark{3},
Larry Bradley\altaffilmark{3},
Marcella Carollo\altaffilmark{4}, 
Tomas Dahlen\altaffilmark{3},
Henry C. Ferguson\altaffilmark{3},
Mauro Giavalisco\altaffilmark{8},
Anton Koekemoer\altaffilmark{3},
Simon Lilly\altaffilmark{4}, 
Ray A. Lucas\altaffilmark{3}, 
Bahram Mobasher\altaffilmark{9},
Nino Panagia\altaffilmark{3,10,11},
Cheryl Pavlovsky\altaffilmark{3} 
}

\altaffiltext{2}{Department of Physics and Astronomy, Johns Hopkins University, Baltimore, MD 21218, USA; sujian@pha.jhu.edu}
\altaffiltext{3}{Space Telescope Science Institute, Baltimore, MD 21218, USA}
\altaffiltext{4}{Institute of Astrophysics, ETH Zurich, CH - 8093 Zurich, Switzerland}
\altaffiltext{5}{UCO/Lick Observatory, University of California, Santa Cruz, CA 95064}
\altaffiltext{6}{Hubble Fellow}
\altaffiltext{7}{University of Colorado, CASA, Dept. of Astrophysical \& Planetary Sciences, 389-UCB,Boulder, CO 80309, USA}
\altaffiltext{8}{Department of Astronomy, University of Massachusetts, Amherst, MA 01003, USA}
\altaffiltext{9}{Department of Physics and Astronomy, University of California, Riverside, CA 92521, USA}
\altaffiltext{10}{INAF- Osservatorio Astrofisico di Catania, Via S. Sofia 78, I-95123 Catania, Italy}
\altaffiltext{11}{Supernova Ltd., OYV 131, Northsound Road, Virgin Gorda, British Virgin Islands}

\begin{abstract}

In this paper, we present a derivation of the rest-frame 1400\AA\ luminosity function (LF) at redshift six from a new application of the maximum likelihood method by exploring the five deepest {\it HST}/ACS fields, i.e., the HUDF, two UDF05 fields, and two GOODS fields. We work on the latest improved data products, which makes our results more robust than those of previous studies. We use un-binned data and thereby make optimal use of the information contained in the dataset. We focus on the analysis to a magnitude limit where the completeness is larger than 50\% to avoid possibly large errors in the faint end slope that are difficult to quantify. We also take into account scattering in and out of the dropout sample due to photometric errors by defining for each object a probability that it belongs to the dropout sample. We find the best fit Schechter parameters to the \six LF are: $\alpha = 1.87 \pm 0.14$, $M_* = -20.25 \pm 0.23$, and $\phi_*=1.77^{+0.62}_{-0.49}\times 10^{-3}$ Mpc$^{-3}$. Such a steep slope suggests that galaxies, especially the faint ones, are possibly the main sources of ionizing photons in the universe at redshift six. We also combine results from all studies at \six to reach an agreement in 95\% confidence level that $-20.45<M_*<-20.05$ and $-1.90<\alpha<-1.55$. The luminosity density has been found not to evolve significantly between \six and $z\sim5$, but considerable evolution is detected from \six to $z\sim3$.

\end{abstract}

\keywords{galaxies: evolution --- galaxies: formation --- galaxies: high-redshift --- galaxies: luminosity function --- methods: data analysis}

\section{INTRODUCTION}

Deep imaging surveys, such as the Great Observatories Origins Deep Survey \citep[GOODS,][]{04giavalisco} and the Hubble Ultra Deep Field \citep[HUDF,][]{06beckwith}, have been extensively analyzed to study galaxy properties out to the reionization epoch. The rest-frame ultraviolet (UV) galaxy luminosity function (LF) is measured for samples of Lyman break galaxies (LBGs) and used  to detect cosmic evolution. The consensus that has developed is that a considerable increase in the space-density of galaxies at the bright end of the LF occurs from redshift \six\citep{04bunker,04yw,06beckwith,06bouwens}\footnote {The results of these groups are summarized in Table \ref{tab:six}.} to $z\sim3$ \citep[e.g.,][]{99steidel}. However, there are still some discrepancies in the interpretation of this evolution, in terms of density, slope, luminosity, or a combination of these. \citet{04bunker} undertake a photometric analysis of the HUDF \idrops and propose that the density increases six-fold from \six to $z\sim3$, in agreement with \citet{06beckwith}. \citet{04yw} push the detection limit deeper to magnitude 30, finding a steeper faint slope at \six compared to $z\sim3$ by 0.2-0.3. Furthermore, \citet{06bouwens} estimate corrections to the measured quantities to account for various observational effects  and conclude that the intrinsic luminosity is $\sim$0.8 mag fainter at $z\sim6$. Their conclusions remain qualitatively unchanged after \citet{09reddy} recently revisit the LF parameters at $z\sim3$. On the other hand, ground-based observations, e.g., \citet{09mclure}, find an even stronger luminosity evolution. 

Different measurements of the luminosity density (LD) or star formation rate (SFR) also give somewhat different results \citep[e.g.,][]{04bunker,06bouwens}. It is important to establish whether these observed differences are due to intrinsic differences in the evolution of different galaxy populations or due to issues with the derivation of the LF.

Spectroscopic confirmations of \six galaxies, e.g., \citet{05malhotra}, \citet{07dow}, \citet{08hathi} and \citet{09vanzella}, have already proven the effectiveness and robustness of the dropout technique in selecting LBGs. However, the faint LBGs, which are essential to determining the faint-end slope of the LF, have not been spectroscopically confirmed because they require impractically long exposure time on large telescopes. 

Therefore, to improve upon the previous studies of the \six LF and to establish its form, a number of difficult issues should be considered. (a) Optimal use of the data: a single field provides us with only a handful of candidates so that some magnitude intervals contain only very few objects. Thus, it is very important to keep all the information. In order to do so, we use un-binned data. (b) Completeness of the catalogs: the correction to the number of objects observed at faint magnitudes is significant due to the detection incompleteness. We adopt a more moderate magnitude limit than other groups in order to avoid  possible uncertainties brought by large corrections. (c) Photometric errors and biases: a strict color cut used for \idrop selection may lose real LBGs and is affected by contaminants. For each galaxy within or outside the selection window, we explicitly consider its probability of being an actual LBG by assuming a Gaussian distribution for the photometric error.

On the basis of the HUDF images \citep[][hereafter paper I]{06beckwith}, the UDF05 images \citep[][hereafter paper II]{07oesch}, and the HUDF09 images \citep{10oesch, 10bouwens}, we are now in the position to study properties of LBGs from $z\sim 0$ to beyond $z\sim8$ utilizing {\em Hubble Space Telescope}'s ({\it HST}) unparalleled deep optical and infrared (IR) view. In this paper, we plan to further develop techniques to derive the LF at \six using the procedures used for $z\sim5$ galaxies in paper II. In particular, we apply the maximum likelihood (ML) method, which is independent of clustering in our sample, to derive the LF and examine whether star forming galaxies, especially the faint ones, are responsible for re-ionizing or keeping the universe ionized at $z\sim6$. 

We adopt $\Lambda$CDM cosmology: $\Omega_M=0.3$, $\Omega_\Lambda=0.7$, and $H_0=70$ km s$^{-1}$ Mpc$^{-1}$. Magnitudes are in the AB system.

\section{DATA}

We work on five {\it HST}/ACS deep fields in four broad bands: F435W ($B_{435}$), F606W ($V_{606}$), F775W ($i_{775}$), and F850LP ($z_{850}$). We use the most recent and updated version of the data, namely: GOODS South (GOODS-S) \& GOODS North (GOODS-N) v2.0 data by \citet{08giavalisco},  the HUDF data from paper I, HUDF NICP12 from paper II, and HUDF NICP34 processed in this work. PyRAF tasks Multidrizzle and Tweakshifts \citep{06koekemoer} help precisely align the images, and SExtractor \citep{96ba} is run in double-image mode with \zs\ as the detection band to generate the catalogs. Our survey covers $\sim$ 350 arcmin$^2$ to a magnitude limit of  $z_{850}\sim29$, identifying $\sim$ 1100 LBG candidates at \six, with an average number of ~350 per realization, as shown in Table \ref{tab:dropas}.

We have not made use of the WFC3/IR data that are becoming available on these fields for two main reasons: (a) The IR data are not available over the full fields, especially only a small portion of the GOODS has been covered. This would force us to reduce the sample size greatly. (b) We think an important component of this work is a comparison with other published results which are based on the simple one color selection rather than a full two-color selection. We do make a quick check of the IR information in the HUDF in Section \ref{sec:s} and will leave a full investigation for future work.

We also do not use ground based data for two main reasons: (a) The two GOODS fields are already large enough to provide good constraints on galaxies brighter than the knee of the Schechter function. (b) We prefer to work with a homogenous data set in terms of filters and detector QE curves.

\section {LUMINOSITY FUNCTION OF LBGs AT \six}

The completeness function $C(m)$ ($m$ is the apparent/detected magnitude) and the selection function $S(m,z)$ ($z$ is the redshift) are measured by performing recovery simulations in the same way as in paper II, i.e., by inserting artificial galaxies into our science images and rerunning SExtractor with the same setup as for the original catalog generation. We use a $\beta-$distribution $-2.2 \pm 0.4$ \citep{05stanway} and a size distribution following a scaling of $(1+z)^{-1}$ as in \citet{04ferguson}. For each redshift bin $\delta z=0.1$, we thus compute the color a galaxy would have with the randomly chosen $\beta-$value and insert it in the images. The input magnitudes are following a flat distribution from $24 - 29$, but the selection function is given at observed magnitudes, simply by computing the fraction of galaxies that we insert with the measured output magnitude which is selected by the \idrop criteria. $C(m)dm$ is defined as the probability that a galaxy of \m in the images is selected in the catalog, which depends strongly on SExtractor parameters such as DEBLEND. Thus, it is important that the recovery simulations are done using the same SExtractor parameters used to derive the catalog. $S(m,z)dmdz$ represents the probability that a LBG at a given redshift \z and at a given observed magnitude \m satisfies the selection criteria. Naturally, the product of these two functions $C(m) S(m, z)dmdz$ is the probability that a galaxy at redshift \z is detected with magnitude \m AND selected as a LBG. 

The UV LF can be expressed in Schechter form as, 
\begin{equation} 
\phi(M)=(0.4\ln 10)\phi_*
10^{0.4(1+\alpha)(M_*-M)}\exp[-10^{0.4(M_*-M)}] 
\end{equation} 
with the absolute magnitude $M=m-DM(z)-K_{cor}(z)$, where $DM(z)$ is the distance modulus and $K_{cor}(z)$ is the
K-correction from observed \zs\ to rest-frame 1400 \AA.

Binned data were initially utilized by many groups to derive the shape of the LF. The observed number of LBGs within the apparent magnitude bin $m_l<m<m_u$ is predicted as
\begin{equation} 
N_i =\int
dz \frac{dV_C}{dz}(z) \int_{m_l}^{m_u} dm C(m) S(m, z)
\phi(M(m,z);\phi_*,M_*,\alpha)\label{eqn:ni} \end{equation} where ${dV_C}/{dz}$ is the comoving volume element of the survey. Binning may lose information, and lead to biased results dependent on the bin size. At the same time, having very few luminous candidates in current high-z surveys, there is uncertainty about the numbers in the bright bins since the candidates could jump into adjacent bins due to photometric errors. Simulations by \citet{08ts} show that binning is likely to affect the confidence regions for the best-fitting parameters. 

To overcome these drawbacks, in this section we present an improved approach based on the ML method \citep[][STY]{22fisher,sty} to make optimal use of every possible LBG in the fields. As also pointed out by \citet{08ts}, the STY ML estimator relies essentially on un-binned data. We determine the shape of the LF by exploring every single detected dropout. First, we find the probability for each galaxy that it could be selected as a LBG, considering the photometric uncertainty of the catalogs (Section \ref{sec:f}). Second, we choose galaxies randomly by the above probability and run our ML process (Section \ref{sec:v}). Third, we repeat the above step enough times to achieve convergence.

\subsection{Selection Criteria} \label{sec:s}

We adopt the \idrop selection criteria from paper I, i.e.,
\begin{eqnarray} 
i_{775}-z_{850} &>& 1.3,\\
S/N(z_{850}) &>& 5, \\
S/N(V_{606})<2 &{\rm or}& V_{606}-z_{850}>2.8.
\end{eqnarray} 
The dominant criterion, i.e., the SExtractor MAG\_ISO color \iz$>1.3$, will be further discussed in Section \ref{sec:f}. The signal-to-noise ratio $S/N(z_{850})>5$ is demanded for each candidate to largely avoid interlopers (later this subsection) or slope steepening (Appendix), and to be consistent in comparing with $z\sim 3$ results from \citet{99steidel} and with $z\sim 5$ results in  paper II. The photometric errors also take into account the correlated errors present in the images as discussed in paper II. In addition, we require for CLASS\_STAR $< 0.75$ if the MAG\_AUTO magnitude \zs$<28.0$ for the HUDF, $<27.5$ for the UDF05, and $<26.5$ for the GOODS in order to remove stellar contamination at the bright end \citep[e.g.,][and paper II]{06bouwens}. Only galaxies with $C(m)>0.5$ have been included to avoid large uncertainty corrections (Table \ref{tab:c}). The selection has been proven to be very efficient and effective. All the spectroscopically confirmed \iz$>1.3$ \six LBGs through the HUDF/GOODS follow-up surveys \citep[e.g.,][]{05malhotra,09vanzella} satisfy our criteria, with only one exception, and no Galactic star could pass the CLASS\_STAR test.

\begin{figure} 
\plotone{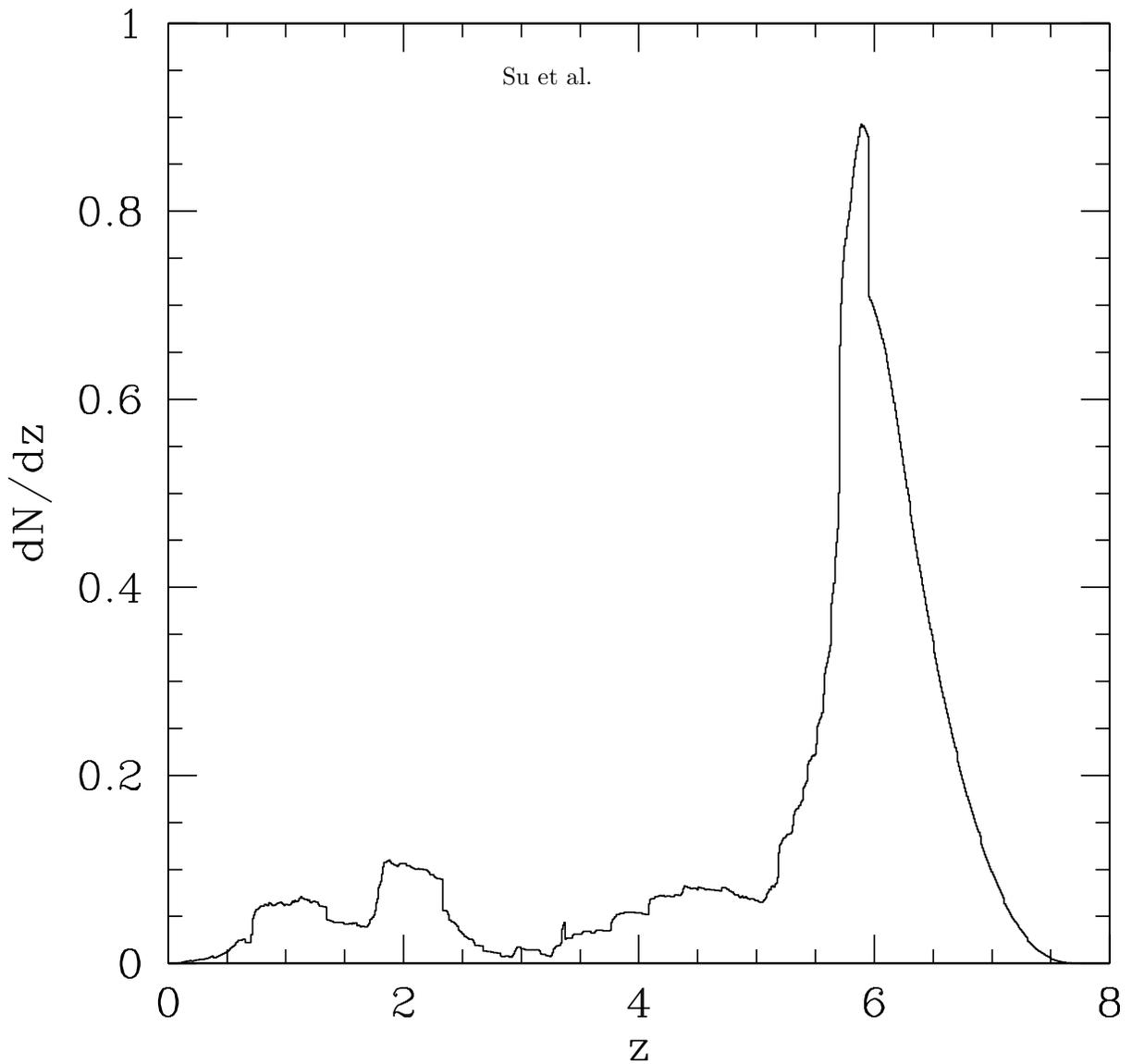} 
\caption{Predicted redshift distribution for \idrops as derived assuming synthetic SEDs and a non-evolving LF in the redshift window $5.7<z<7$. The total interloper fraction is estimated to be 24\% and is primarily contributed by lower redshift galaxies selected as LBGs due to the aliasing between the Lyman break and the 4000\AA\ break. The model is pessimistic and at the relatively bright end (\zs$<27.5$) comparison with \citet{05malhotra} shows a factor of two fewer interlopers than predicted by the model.
}\label{f1} 
\end{figure}

\begin{figure} 
\plotone{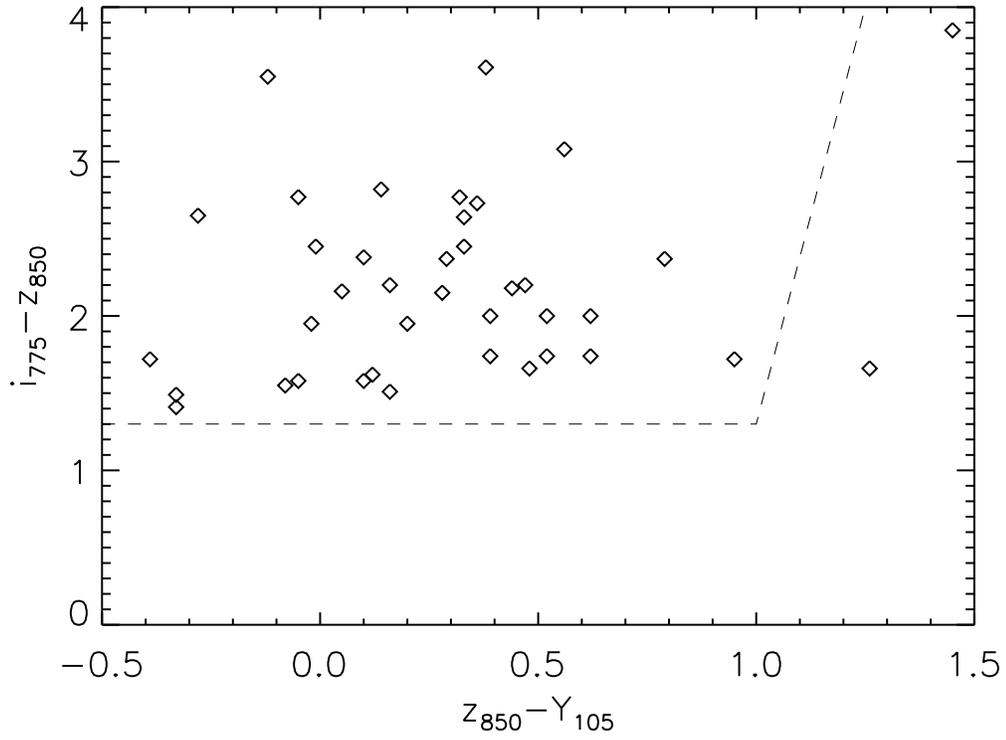} 
\caption{Color-color diagram of the HUDF \iz$>1.3$ candidates. \zs \ image is rescaled to match Y$_{105}$ to get the \zs-Y$_{105}$ color. The dash lines give a possible \idrop selection criterion, namely \iz$>1.3$ and \zs-Y$_{105}$ $<$ 1 + 0.09(\iz-1.3). }\label{f22} 
\end{figure}

We have estimated the possible fraction of interlopers by applying our selection criteria of equations (3)-(5) to a library of $\sim$3000 synthetic SEDs built on Bruzual-Charlot  models \citep{03bc}, adopting the LF derived by \citet{99steidel} at $z\sim3$ and no evolution. The models include the effects of intergalactic absorption \citep{95madau}, and span a wide range of metallicities (0.04-2.5 $Z_\sun$), dust reddening (no extinction, or eight values logarithmically spaced between $A_V = 0.05$ and 6.4), emission lines (no lines or lines computed from first principles from UV SED for Hydrogen and fit to Cloudy models for metal lines), and different star formation histories (burst, constant at fixed metallicity, constant at evolving metallicity without infall or with infall, two component models with an old component and a young one which may include emission lines). We consider 18 ages logarithmically spaced between 1 Myrs and 18 Gyrs. And no model older than the universe is included. We can see from the resulting redshift distribution (Fig. \ref{f1}) that there is a lower redshift population $z\sim1,2$ of galaxies that may be selected as LBGs at \six due to the aliasing between the Lyman break and the 4000\AA\ break \citep[See e.g.,][for more discussions]{10dahlen}. In Fig. \ref{f22}, we have identified our \iz$>1.3$ candidates detected by the WFC3 F105W (Y$_{105}$) band in the HUDF to verify that our sample does not have many interlopers.

\subsection{\f}\label{sec:f} 
Photometric scatter introduces large uncertainties in numbers and magnitudes of the LBG candidates, and therefore, in determined properties of the LF. If a strict color cut such as \iz$>1.3$ was applied, the impact of photometric errors would not be fully explored, and many real LBGs with a little bluer measured color may be missed due to photometric errors. A relaxed cut, e.g., \iz$>0.9$, on the other hand, suffers from larger contaminations. For example, \citet{05malhotra} found five objects at intermediate redshifts and four intrinsic \six galaxies within $0.9<$\iz$<1.3$, which means the contamination rate in the relaxed color window may be as high as $5/(5+4)=56\%$.

To account for this effect, we calculate the probability that each object is an LBG, which decides how often it could contribute to the later ML process. If $p(m)dm$ is the probability that a galaxy is of magnitude \m in the catalog, then the \f of \iz$>1.3$ LBG candidates is defined as: 
\begin{equation} 
f=\int \mathrm{d} z_{850} \mathrm{d} i_{775} \ p(z_{850}) p(i_{775})
\end{equation} 
where the integration of \is\ is taken over \iz$>1.3$. The real magnitude \m is assumed to be a Gaussian distribution around its cataloged magnitude $m_c$ (See Appendix for details). In practice, one could find the values of \f with a Monte Carlo method by simply generating Gaussian distributed magnitudes repeatedly to see how often the \iz$>1.3$ color would be satisfied. A 2-$\sigma$ magnitude limit is adopted if there is no detection in the \is-band. 

\begin{figure} 
\plotone{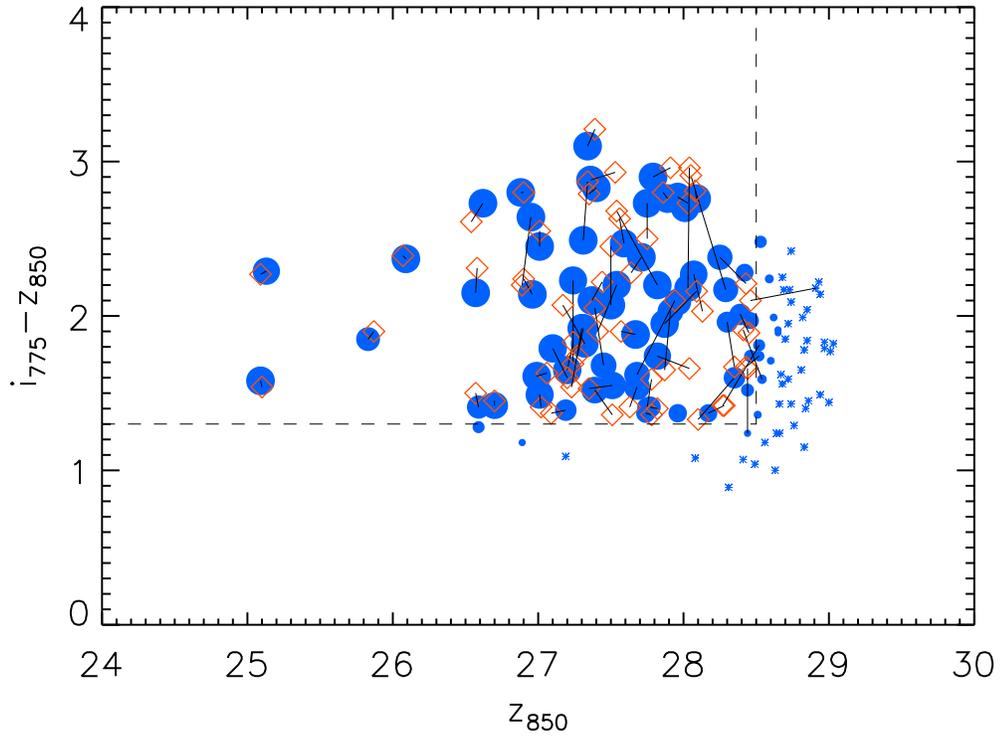} 
\caption{Color-magnitude diagram of the HUDF candidate pool. Candidates are indicated as filled circles whose radius is proportional to the $f$-factor, and asterisks are those objects with \f less than 0.2. The diamonds are \idrops selected in one realization for use in later ML process, and the line segments connect the cataloged and realized positions in the diagram.}\label{f2} 
\end{figure}

It is easy to see that $f>0.5$ when the cataloged \iz$>1.3$ while $f<0.5$ when the cataloged \iz$<1.3$, and $f=0.01$ corresponds to the cataloged \iz\ $\sim$ 0.9 when the \zs\ and \is\ errors are both 0.2. All $f\geq0.01$ galaxies are used in the subsequent ML analysis, i.e., 1\% chance of being included in one realization. Table \ref{tab:dropas} shows that essentially about 25\% - 50\% candidates in each field will participate in one realization, which brings our sample in agreement with other groups within the magnitude window in study, such as \citet{07bouwens}. (See Fig. \ref{f2} and Table \ref{tab:udf}.)

\subsection{V-Matrix}\label{sec:v}

Due to the unique long tail of the ACS \zs-filter, the K-correction can be as large as 2.2 mag at $z=5.7$ and goes down to 0.3 mag at $z=7.0$. Thus, with distance modulus varying by 0.5 mag there could be a 2.4-mag scatter in UV rest frame absolute magnitudes in realizations at $5.7<z<7$ for any given observed \zs-magnitude. In other words, the relation between \M and \m is very uncertain. Therefore, applicable at where \M is relatively insensitive to redshift or the redshift span is relatively small, the effective volume $V_{eff}$ technique does not fit in our case. This forces us to seek a new formalism.

We define the apparent LF as 
\begin{equation} 
\Phi(m) =\frac{C(m)}{V_{eff}} \int dz S(m, z) \frac{dV_C}{dz}(z) \phi(M;m,z) \label{eqn:phi}
\end{equation} 
and it does not need to be of Schechter form. The V-matrix is therefore,
\begin{equation} 
V(m,z) \equiv  C(m) S(m, z) \frac{dV_C}{dz}(z) 
\end{equation} 
and $V_{eff}(m)=\int dz V(m,z)$. We maximize the likelihood function $\ln L= \sum_i \ln p(m_i)$ where 
\begin{eqnarray}
&&p(m_i) = \Phi(m_i)/\int dm\Phi(m)\\
&=& \int dz V(m_i,z) \phi (M_i)/\int \int dm dz V(m,z) \phi(M) \nonumber
\end{eqnarray} 
The integrations are always taken over the region of interest, for example for the HUDF, $5.7<z<7.0$ and $24.0<m<28.5$. (The bright limit is introduced for calculations only when there is no candidate detected beyond this magnitude, and an even brighter limit will not affect the results since the LF is greatly suppressed at this end.) $C(m)$ has been included in the calculation of $V(m,z)$ so that there is no additional completeness correction factor in $p(m_i)$. \footnote {We note that \citet{85marshall} adopted a similar approach to ours and he did not have to take the integration of redshift as shown above since the redshifts of their objects were already known.}

When combining different fields, e.g., the GOODS and the HUDF, no additional rescaling factor is needed in the ML method \citep{08ts}. The inputs to the ML process are the V-matrix and the magnitudes \m of selected candidates. In each realization, candidates are selected from the pool in a probability as to their $f$-factor. The outputs are $M_*$ and $\alpha$ in as many as possible realizations, when the averages and errors have been convergent. The uncertainty of \m considered in the ML process only yields minor errors when several hundreds of galaxies are surveyed (Appendix). $\phi_*$ is determined by $\chi^2$ fit to the observed LBG densities with respect to the 1-$\sigma$ 2-parameter contour of $M_*$ and $\alpha$. 

\begin{figure} 
\plotone{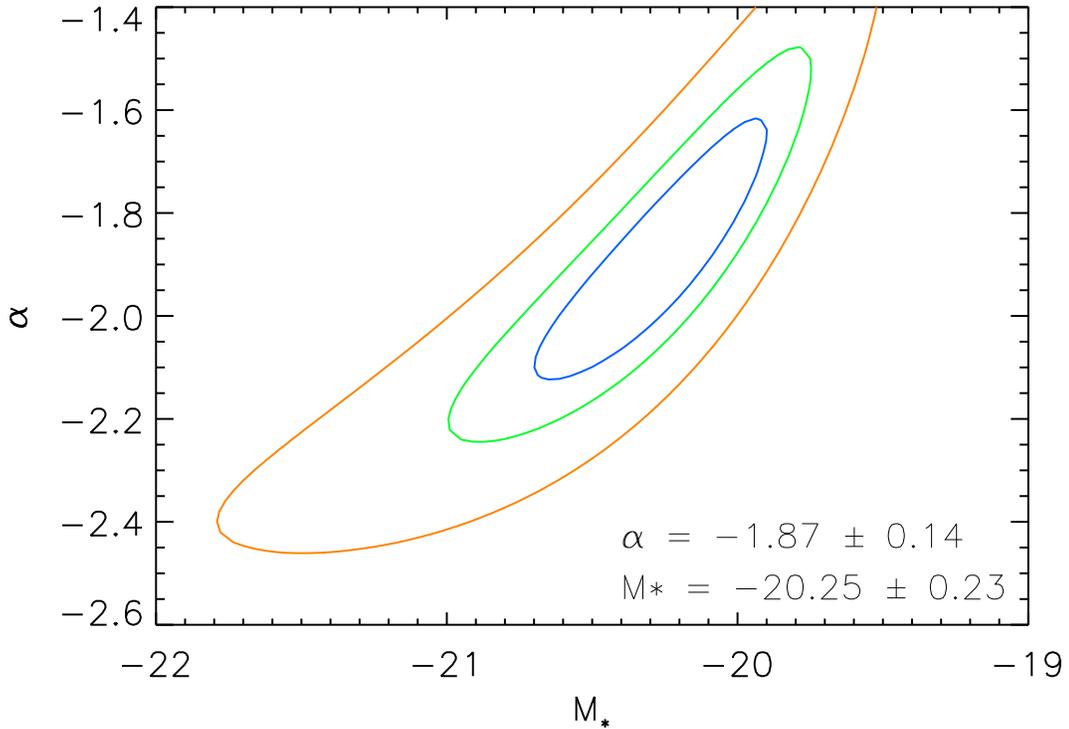} 
\caption{Likelihood contour for the best fit Schechter parameters of the \six LF. The contours, inner to outer, stand for 1-parameter 1-$\sigma$, 2-parameter 1-$\sigma$, and 1-parameter 2-$\sigma$ likelihood contours averaged over realizations for use in the ML process.}\label{f3} 
\end{figure}

The LF parameters we derive for \six are: $\alpha=-1.87\pm0.14$, $M_*=-20.25\pm0.23$, and $\phi_*=1.77^{+0.62}_{-0.49}\times 10^{-3}$ Mpc$^{-3}$, as illustrated in Fig. \ref{f3}. We notice our faint end slope $\alpha$ is slightly steeper than that from some other studies. This could partly be caused by a steeper slope at $z\sim7$ \citep[e.g.,][]{10oesch,10trenti} since we include up to $z=7$ LBGs in our estimate of the \six LF.

\subsection{Evolution of $\phi_*$} 
Since we are investigating a relatively large redshift range $5.7<z<7.0$ and finding indication of LF evolution, it is a good sanity check for us to explicitly consider the effect of evolving LF parameters. Assuming $M_*$ and $\alpha$ are uniform in this redshift range, we assign a linear evolution $\phi_*(z)=\phi_*(6.3)[1-\frac{5}{7}(z-6.3)]$ and repeat the analysis described in Section \ref{sec:s} - \ref{sec:v}. We find that $\alpha=-1.92\pm0.13$, $M_*=-20.22\pm0.21$. The closeness to our derived parameters for no evolution, i.e., $\alpha=-1.87\pm0.14$ and $M_*=-20.25\pm0.23$, shows that our results are robust with respect to an evolution of the LF normalization within the redshift range of \idrops.

\subsection{Evolution of $M_*$} 
Similar considerations to those in the previous subsection lead us to explore a variation of $M_*$ within the  \idrop\  redshift window. We do so by assigning $\M_*(z)=\M_*(5.9)+0.36(z-5.9)]$ \citep{07bouwens,10oesch} while keeping uniform values of $\alpha$ and $\phi_*$. We find $\alpha=-1.91\pm0.08$, which is also within one sigma of our non-evolving derivation.

\section{COMPARISON TO OTHER RESULTS} 
We have verified the internal consistency and robustness of our results and we are now ready to compare them to other studies.

\subsection{Most Probable \six LF} 
\begin{figure} 
\plotone{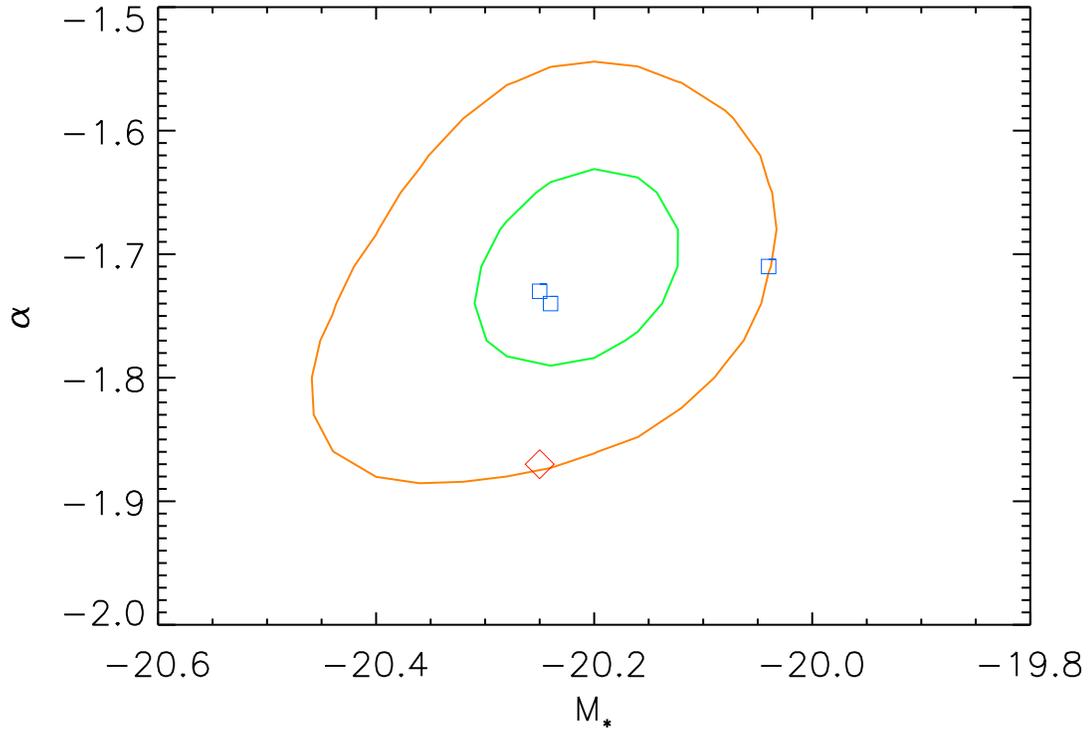} 
\caption{Most probable parameter space at \six based on ten studies. The inner contour includes 68\% probability and the outer 95\%, assuming all the studies are independent and correct. Two nearby squares are from Bouwens et al. (2006, 2007), a third square is from \citet{09mclure} who combine their data with \citet{07bouwens}, and the diamond is from this work. As illustrated in Fig. \ref{f3}, $M_*$ and $\alpha$ are strongly correlated, so we do not plot their error bars, which can be found in Table \ref{tab:six}.}\label{f4} 
\end{figure}

To deal with the weighted average of results from different groups, we follow \citet{96press}. The probability of getting observed variable(s) $H_0$ from data or measurements $D$ is 
\begin{equation} 
P(H_0|D)\propto \displaystyle\prod_i
  (P_{Gi}+P_{Bi}) \label{eqn:GB} 
\end{equation} 
Here $P_{Gi} \sim\frac{1}{\sigma_i}\exp [\frac{-(H_i-H_0)^2}{2\sigma_i^2}]$ and $P_{Bi} \sim \frac{1}{S}\exp [\frac{-(H_i-H_0)^2}{2S^2}]$ are the probability distributions of ``good" and ``bad" measurements, respectively, where $i$ denotes different measurements, and $S$ should be assigned to be large enough to ensure that measurements do not conflict with each other. When extending this method to two-dimensional analysis, we also consider the correlation between $M_*$ and $\alpha$ (Fig. \ref{f3}). \citet{96press} puts almost no weight on those measurements without errors where $P_{Gi}=0$ and $P_{Bi}$ is widely spread. Instead, we assume a moderate error of 0.3 for those six groups, i.e., \citet{04bouwens}, \citet{04bunker}, \citet{04dickinson}, \citet{04yw}, \citet{05malhotra}, and paper I. Combined with the other four measurements providing errors, i.e., \citet{06bouwens}, \citet{07bouwens}, \citet{09mclure}, and this work, we find there is about a 95\% chance that $-20.45<M_*<-20.05$ and $-1.90<\alpha<-1.55$, assuming all the current studies are independent and correct.  (See Table \ref{tab:six} and Fig. \ref{f4}.) 

\subsection{$z\sim5$ LBGs LF revisited} 
In order to further test the method used here, we derive the faint end slope of the $z\sim5$ LBG LF using the same catalogs and the same selection criteria as those in paper II. To study the HUDF and NICP12 data that lack enough bright candidates to determine $M_*$, we fix $M_*=-20.7$ to find $\alpha=-1.72\pm0.04$, which is in agreement with the previous results (their Table 3). Thus, our method, designed to deal with the varying K-correction in \zs\ and to account for additional uncertainties, is equivalent to our previous method in the simpler \vdrop case.

\subsection{$z=3\sim6$ Luminosity Density} 
\begin{figure} 
\plotone{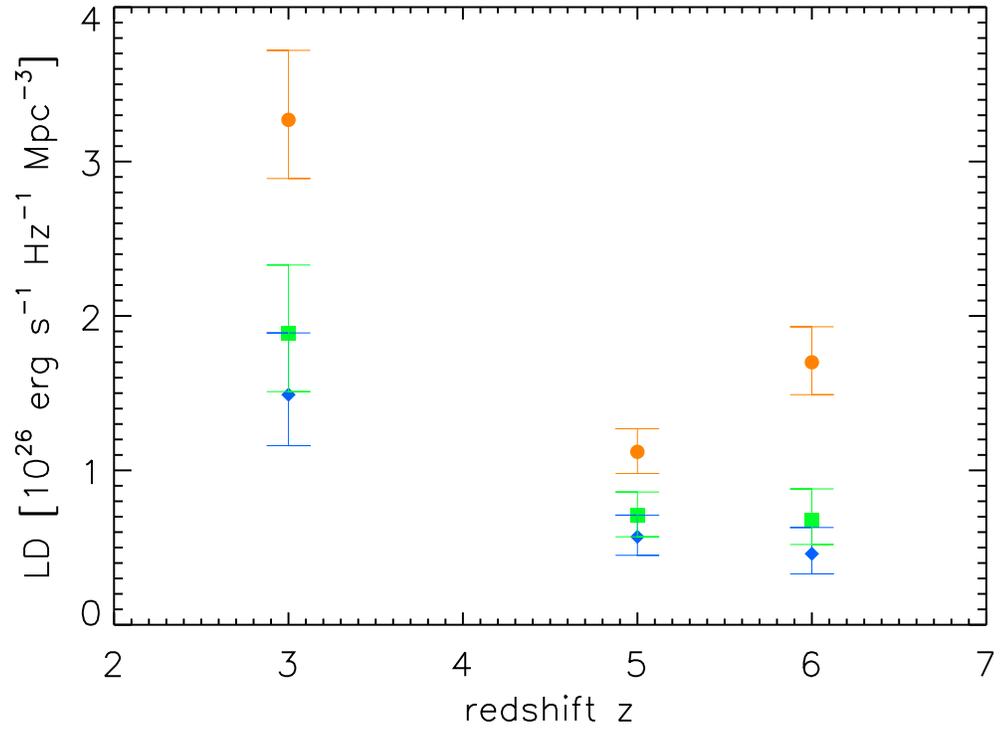} 
\caption{Luminosity density from $z\sim3$ to \six integrated to 0.3 (diamonds), 0.2 (squares), 0.04 (circles) of $L_*(z=3)$. $z\sim3$ data calculated from \citet{09reddy}, $z\sim5$ data calculated from paper II, and \six data calculated from this work. See Table \ref{tab:ld} for the numbers.}\label{f6} 
\end{figure}

The luminosity density (LD) at redshift \z equals $\int L\phi(L) dL = L_*(z) \phi_*(z) \int_{x_0}^\infty x^{1+\alpha} e^{-x} dx$, where $x=L/L_*(z)$. We find there is considerable evolution between \six and $z\sim 3$, but no statistically significant evolution between \six and $z\sim 5$. More details are in Table \ref{tab:ld} and Fig. \ref{f6} where $x_0=aL_*(3)/L_*(z)$ and a=0.3,0.2,0.04. At lower redshifts there are fewer recombinations in the diffuse medium and therefore the required flux density to keep the universe ionized increases with increasing redshift. If the universe has finished reionizing at \six, then it will be kept ionized at $z\sim5$ since the required LD at $z\sim5$ is less than that at \six and the observed ones are close to each other.

\section{CONCLUSIONS} In this paper, we have reported the results of a study of a large sample of faint LBGs in the redshift interval $5.7<z<7.0$. Working on the five deepest {\it HST} fields with their most updated data, we account for the effect of photometric errors by introducing the factor f as the probability of each galaxy to be an LBG. We employ un-binned data to keep all the information and to avoid bias, and we develop a modified ML process to reduce the effect of the uncertain relation between M and m. Our best-fitting Schechter function parameters of the rest-frame 1400\AA\ LF at redshift \six are: $\alpha=-1.87\pm0.14$, $M_*=-20.25\pm0.23$, and $\phi_*=1.77^{+0.62}_{-0.49}\times 10^{-3}$ Mpc$^{-3}$, which suggest evolution of $M_*$, possible steepening of $\alpha$, and no change of $\phi_*$ compared to their values at $z\sim3$. Such a steep slope suggests that galaxies, especially the faint ones, are possibly the main sources of ionizing photons in the universe at redshift six \citep{04stiavelli}. Combining ten previous studies at \six with the extended Press method, we find that the most probable LF favors $-20.45<M_*<-20.05$ and $-1.90<\alpha<-1.55$ at the 95\% confidence level. The LD has been found not to evolve significantly between \six and $z\sim5$, but considerable change is detected from \six to $z\sim3$.

If $\alpha$ remains constant from \six to $z\sim3$ as stated by e.g., \citet{07bouwens} and \citet{09reddy}, it will be difficult to tell the intrinsically evolving parameter, $M_*$ or $\phi_*$, from faint LBGs only, while too few bright LBGs are found due to the limited area of current deep surveys. Ground-based surveys such as the Subaru Deep Field \citep{05shi,09mclure} are extremely efficient in detecting bright LBGs in a large field of view and might clarify whether $M_*$ or $\phi_*$ alone is not responsible for the change of LF, while splitting the \zs-band into two separate bands may be useful to isolate the effect of a possible slope steepening \citep{05shi}. We look forward to including IR data from WFC3 on board {\it HST} to improve the selection of \six LBG candidates, and the bright end of the LF will be better determined when the data from CANDELS/ERS \citep[e.g.,][]{10bbouwens} and the BoRG survey \citep{11trenti} are becoming available.

\acknowledgments
JS and MS have been partially supported by NASA grant NAG 5-12458. PO is supported by NASA through Hubble Fellowship grant HF-51278.01. Support for program \#10632 and \#11563 was provided by NASA through a grant from the Space Telescope Science Institute, which is operated by the Association of Universities for Research in Astronomy, Inc., under NASA contract NAS 5-26555.

\appendix 
\section{More About Photometric Scatter and Flux Boosting} 
We assume the photometric scatter is in a Gaussian distribution, thus the probability of a galaxy arriving on the detector as magnitude $m$ but cataloged in $m'$ equals 
\begin{eqnarray} 
G(m,m',\sigma) = \frac{1}{\sqrt{2\pi\sigma^2}}\exp[-\frac{(m-m')^2}{2\sigma^2}] 
\end{eqnarray} 
and the measured LF will be 
\begin{eqnarray} 
\phi'(m) &=& \int \phi(m')G(m,m',\sigma)dm'
\end{eqnarray} 
where $\phi$ is the actual LF, i.e., Equation (7) in Section \ref{sec:v} .

When the photometric error $\sigma$ is very small, $G$ takes the limit of the Dirac function and it is always true $\phi'\equiv\phi$. When the surveys are pushed close to the detection limit, $\sigma$ is not negligible and also far from uniform in the magnitude window. To satisfy $S/N=10$ at $m=m_*$ and $S/N=5$ at $m=m_*+2.5$, a guess would be 
\begin{eqnarray} 
\sigma(m) = \frac{2.5}{\ln 10} \frac{1}{2(m_*-m)+10} 
\end{eqnarray} 
Simulations show that the effect of flux boosting from fainter magnitudes outside our selection window is negligible. But as shown in Table \ref{tab:boosting}, if $\sigma(m)$ increases much faster with $m$, or if lower S/N candidates are included, there will be considerable steepening at the faint end due to the photometric scattering. We simulate 4000 objects according to the given LF parameters, i.e., $m_*$ is fixed and $\alpha$ =  -1.7 in [$m_*$-3.5,$m_*$+4.5]. Their magnitude errors are assumed to be in the form of $10^{0.3(m-m_*)}$ which comes from the real data of the HUDF. For each realization, the change of magnitudes brought by their errors will also change their detected S/N. We choose those with the S/N$>$ 5 and lying within [$m_*$-2.5,$m_*$+2.5] to determine the slope. This process repeats for different combinations of S/N$>$ 5, 7, 9 and $\alpha$ =  -1.5, -1.7, -1.9. We can see from Table \ref{tab:boosting} that if the S/N is kept $>$ 5, the steepening of the faint end slope by the flux boosting is less than 0.1.

\begin{figure} 
\plotone{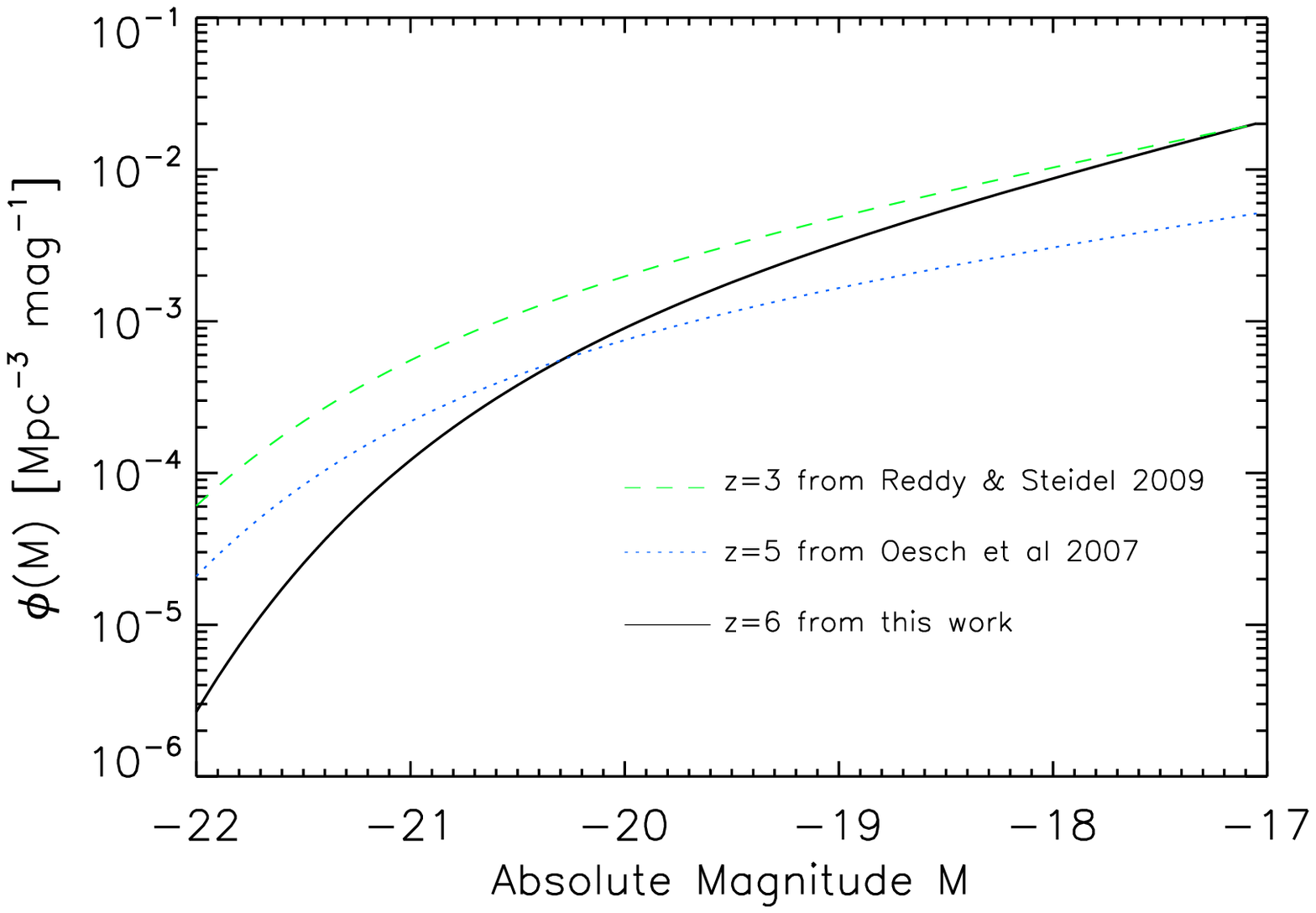} \caption{Luminosity function from $z\sim3$ to \six. }\label{f5} 
\end{figure}

\begin{deluxetable}{cccccc} 
\tablecaption{Dropouts in Our Sample\label{tab:dropas}} 
\tablewidth{0 pt} \tablecolumns{6} 
\tablehead{\colhead{} &
\colhead{HUDF} & \colhead{GOODS-S} & \colhead{GOODS-N} & \colhead{NICP12}& \colhead{NICP34} } 
\startdata 
$N_{tot}$\tablenotemark{a}&	 115&	373&	502&	120&	54\\
$N_s$\tablenotemark{b}&	58.1$\pm$2.3&	103.1$\pm$6.5&	116.2$\pm$7.0&	33.9$\pm$3.0&	23.5$\pm$5.7
\enddata 
\tablenotetext{a} {Total number of galaxies in our candidates pool.} 
\tablenotetext{b} {Average number of galaxies in one realization.} 
\end{deluxetable}

\begin{deluxetable}{ccccc} 
\tablecaption{Completeness of the Fields \label{tab:c}}
\tablewidth{0 pt} \tablecolumns{5} \tablehead{\colhead{$z_{850}$\tablenotemark{a}} & \colhead{HUDF} & \colhead{NICP12} & \colhead{NICP34} & \colhead{GOODS}} \startdata
24.25&	0.95&	0.95&	0.96&	0.96\\ 24.75&	0.94&	0.94&	0.95&	0.96\\
25.25&	0.93&	0.94&	0.94&	0.95\\ 25.75&	0.92&	0.93&	0.93&	0.94\\
26.25&	0.91&	0.92&	0.92&	0.86\\ 26.75&	0.89&	0.91&	0.87&	0.61\\
\cline{5-5} 27.25&	0.87&	0.86&	0.70&	0.30\\ \cline{4-4}
27.75&	0.79&	0.72&	0.43&	0.10\\ \cline{3-3}
28.25\tablenotemark{b}&	0.60&	0.47&	0.19&	...\\ \cline{2-2}
28.75&	0.37&	0.23&	0.07&	... 
\enddata 
\tablenotetext{a} {Central bin magnitude.}
\tablenotetext{b} {Only data with completeness above half are considered to avoid large uncertainty corrections.}
\end{deluxetable}

\begin{deluxetable}{cccc} 
\tablecaption{Binned numbers of \idrops in the HUDF\tablenotemark{a}
\label{tab:udf}} \tablewidth{0 pt} \tablecolumns{4} \tablehead{\colhead{$z_{850}$\tablenotemark{(1)}} &
\colhead{N$_c$\tablenotemark{(2)}} & \colhead{N$_f$\tablenotemark{(3)}} &
\colhead{N$_s$\tablenotemark{(4)}}} \startdata 24.75&	0&	0.00&	0.01$\pm$0.12\\
25.25&	2&	1.95&	1.94$\pm$0.25\\ 25.75&	1&	0.83&	0.84$\pm$0.39\\
26.25&	1&	1.00&	1.26$\pm$0.50\\ 26.75&	8&	8.42&	8.11$\pm$1.47\\
27.25&	16&	15.52&	16.54$\pm$1.98\\ 27.75&	21&	19.22&	18.07$\pm$2.34\\
28.25&	14&	11.13&	11.30$\pm$2.19 
\enddata 
\tablenotetext{a} {for illustration only, not for later calculations.} 
\tablenotetext{(1)} {Central bin magnitude.} 
\tablenotetext{(2)} {Number of \iz$>1.3$ \idrops from the catalog without corrections.} 
\tablenotetext{(3)} {Number of \idrops weighted with their $f$-factor.} 
\tablenotetext{(4)} {Number of \idrops in simulations considering $f$-factor.}
\end{deluxetable}

\begin{deluxetable}{lcccc} 
\tablecaption{Studies of the \six Luminosity Function\label{tab:six}} 
\tablewidth{0 pt} \tablecolumns{5} 
\tablehead{\colhead{References} & \colhead{Fields\tablenotemark{a}} & \colhead {N\tablenotemark{b}} & \colhead{$\alpha$} & \colhead{$M_*$}} 
\startdata 
\citet{04bouwens}& UDF PFs (28.1)&	30&	-1.15 & -20.26 \\ 
\citet{04bunker}& HUDF (28.5)&	54&	$\leqslant$-1.60 & -20.87 \\ 
\citet{04dickinson}& GOODS (26.0)&	5&	-1.60 (fixed) & -19.87 \\ 
\citet{04yw}& HUDF (30.0)&	108&	(-1.90,-1.80)\tablenotemark{c} & -21.03 \\ 
\citet{05malhotra}& HUDF (27.5)&	23\tablenotemark{d}&	-1.80 (fixed) & -20.83 \\ 
paper I& HUDF (29.0)&	54&	-1.60 (fixed) & -20.5 \\
\citet{06bouwens}& HUDF (29.2) ...\tablenotemark{e}&	506&	-1.73$\pm$0.21 & -20.25$\pm$0.20 \\ 
\citet{07bouwens}& HUDF (29.3) ...\tablenotemark{f}&	627&-1.74$\pm$0.16 & -20.24$\pm$0.19 \\ 
\citet{09mclure}& UDS (26.0)&	157\tablenotemark{g}&	-1.71$\pm$0.11 & -20.04$\pm$0.12 \\ 
this work& HUDF (28.5) ...\tablenotemark{h}&	1164&	-1.87$\pm$0.14 & -20.25$\pm$0.23 
\enddata 
\tablenotetext{a} {The fields and $z_{850}$-band detection limit studied by the reference.} 
\tablenotetext{b} {The number of candidates.} 
\tablenotetext{c} {$-1.9<\alpha<-1.8$.} 
\tablenotetext{d} {all spectroscopically confirmed.} 
\tablenotetext{e} {HUDF (29.2)+HUDF-Ps (28.5)+GOODS (27.5).} 
\tablenotetext{f} {HUDF (29.3)+HUDF05 (28.9)+HUDF-Ps (28.6)+GOODS (27.6).}
\tablenotetext{g} {plus binned data points from \citet{07bouwens}.} 
\tablenotetext{h} {HUDF (28.5)+UDF05 (28.0)+GOODS (27.5)} 
\end{deluxetable}

\begin{deluxetable}{cccc} 
\tablecaption{Evolution of the Luminosity Density\tablenotemark{a}\label{tab:ld}} 
\tablewidth{0 pt}
\tablecolumns{4} \tablehead{\colhead{ }& \colhead{\citet{09reddy}}&\colhead{paper II} & \colhead{this work}\\&($z\sim3$)&($z\sim5$)&(\six)} 
\startdata
$M_*$			&-20.97$\pm$0.14			&	-20.78$\pm$0.21	& -20.25$\pm$0.23\\
$\alpha$		&-1.73$\pm$0.13				&	-1.54$\pm$0.10	& -1.87$\pm$0.14\\
$\phi_*$\tablenotemark{b}	&1.71$\pm$0.53		&	0.9$^{+0.3}_{-0.3}$	&	1.77$\err{0.62}{0.49}$\\
$L_*$\tablenotemark{c}		&1.06$\err{0.15}{0.13}$		&	0.89$\err{0.19}{0.16}$	&	0.55$\err{0.13}{0.11}$\\
LD0.3\tablenotemark{d}		&1.49$\err{0.40}{0.33}$	&	0.57$\err{0.14}{0.12}	$&	0.46$\err{0.17}{0.13}$\\
LD0.2		&1.89$\err{0.44}{0.38}$	&	0.71$\err{0.15}{0.14}$	&0.68$\err{0.20}{0.16}$\\
LD0.04		&3.27$\err{0.45}{0.38}$	&	1.12$\err{0.15}{0.14}$	&1.70$\err{0.23}{0.21}$ 
\enddata 
\tablenotetext{a} {See Fig. \ref{f6} for the graph.} 
\tablenotetext{b} {in units of $10^{-3}$ Mpc$^{-3}$.} 
\tablenotetext{c} {in units of $10^{29}$ erg s$^{-1}$ Hz$^{-1}$.} 
\tablenotetext{d} {in units of $10^{26}$ erg s$^{-1}$ Hz$^{-1}$ Mpc$^{-3}$. LD0.3 means that the LD is integrated from $0.3L_*(z=3)/L_*(z)$.}
\end{deluxetable}

\begin{deluxetable}{cccc} 
\tablecaption{The Steepening of $\alpha$ by the flux boosting for $\sigma(m)\propto10^{0.3(m-m_*)}$ \label{tab:boosting}} 
\tablewidth{0 pt}
\tablecolumns{4} \tablehead{\colhead{ }& \colhead{$\alpha=-1.5$}&\colhead{$\alpha=-1.7$} & \colhead{$\alpha=-1.9$}}
\startdata
$S/N>3$&0.14&0.18&0.26\\
$S/N>5$&0.03&0.04&0.08\\
$S/N>7$&0.01&0.02&0.03
\enddata 
\end{deluxetable}

\end{document}